\documentclass[11pt]{article}

\usepackage{graphicx}
\usepackage{epstopdf}
\usepackage{amsmath}
\usepackage{amssymb}
\usepackage{amsfonts}
\usepackage{amsthm}
\usepackage[usenames]{color}
\usepackage{array}

\usepackage[
      colorlinks=true,
      linkcolor=blue,
      urlcolor=blue,
      filecolor=blue,
      citecolor=red,
      pdfstartview=FitV,
      pdftitle={},
      pdfauthor={},
      pdfsubject={},
      pdfkeywords={},
      pdfpagemode=None,
      bookmarksopen=true
]{hyperref}

\usepackage{epsfig}

\usepackage{hyperref}

\renewcommand{\(}{\left(}
\renewcommand{\)}{\right)}
\newcommand{\p}{\partial}
\newcommand{\bea}{\begin{eqnarray}}
\newcommand{\eea}{\end{eqnarray}}
\newcommand{\ba}{\begin{array}}
\newcommand{\ea}{\end{array}}
\newcommand{\ee}{\end{equation}}

\numberwithin{equation}{section}

\begin{document}

\begin{flushright}
\texttt{\today}
\end{flushright}

\begin{centering}

\vspace{2cm}

\textbf{\Large{
Flat-Space Holography \\ and  Stress Tensor of Kerr Black Hole  }}

  \vspace{0.8cm}

  {\large Omid Baghchesaraei$^a$,  Reza Fareghbal$^a$, Yousef Izadi$^b$, }

  \vspace{0.5cm}

\begin{minipage}{.9\textwidth}\small
\begin{center}

{\it $^a$ Department of Physics, 
Shahid Beheshti University, 
G.C., Evin, Tehran 19839, Iran.  }\\

{\it $^b$ Department of Physics, 
Florida Atlantic University, 
Boca Raton, FL 33431 USA.  }\\

  \vspace{0.5cm}
{\tt omidbaghchesaraei@gmail.com, r$\_$fareghbal@sbu.ac.ir, yizadi2015@fau.edu}
\\ $ \, $ \\

\end{center}
\end{minipage}


\begin{abstract}
 We propose a stress tensor for the Kerr black hole written in the Boyer-Lindquist coordinate. To achieve this, we use the dictionary of the Flat/CCFT correspondence  and take the flat-space limit from the quasi-local stress tensor of the four-dimensional Kerr-AdS black hole.  The proposed  stress tensor  yields  the correct values for the mass and angular momentum of the Kerr black hole at spatial infinity. We also calculate some components of the energy momentum tensor of the three dimensional CCFT and show that they are consistent with the holographic calculation of the Kerr black hole. The calculation we present in this paper is another confirmation for the Flat/CCFT proposal. 
\end{abstract}

\end{centering}

\newpage



\section{Introduction}
 Taking the flat-space limit from the calculations of gravity side in the AdS/CFT correspondence can be used to make some progress in the flat-space holography. The main question is the corresponding operation of the flat-space limit in the boundary theory. There is a proposal in \cite{Bagchi:2010zz},\cite{Bagchi:2012cy} that connects the flat-space limit of asymptotically AdS spacetimes in the bulk side to the ultra-relativistic contraction of the boundary CFT. This proposal is based on the observation  that asymptotic symmetry of asymptotic flat spacetimes at null infinity (which is known as BMS symmetry\cite{BMS}-\cite{aspects}) is isomorphic to an Inonu-Wigner contraction of conformal symmetry. 
 
 It is known that the original coordinates in which the asymtotically AdS spacetimes are written are important in taking the flat-space limit. In three dimensions, one can use a BMS coordinate for writing generic asymptotically AdS spacetimes and take the flat-space limit \cite{Barnich:2012aw}. The final spacetimes are appropriate for studying problems at null infinity. Using these spacetimes, it has been shown in \cite{Barnich:2012aw} that the BMS$_3$ symmetry which is the asymptotic  symmetry at null infinity of three dimensional spacetimes is given by taking suitable limit from the conformal symmetry. 
 
  One may expect the same story in four dimensions and start with a generic solution of asymptotically AdS spacetimes in four dimensions written in the BMS gauge. One would then take the flat space limit and find the generic asymptotically flat spacetimes. As it was reviewed in paper \cite{Lambert:2014poa}, the angular part of metric in the asymptotically flat spacetimes, expanded in powers of radial coordinate, has some terms which are absent for the asymptotically AdS case (see formula (2.54) and (2.103) of \cite{Lambert:2014poa}). Thus, it seems that the BMS gauge is not appropriate in four dimensions  when one wants to study asymptotically flat spacetimes by taking flat space limit from the asymptotically AdS spacetimes.  
  However this is not the end of story and it is straightforward to check that the flat-space limit of the Kerr-AdS black hole written in the Boyer-Lindquist coordinates yields the Kerr solution in the same coordinate. Therefore we expect that holographic calculations for the Kerr-AdS black hole in the context of AdS/CFT correspondence can be well-defined in the flat-space limit when one uses the proposal in \cite{Bagchi:2012cy}. 
 
 However,  the radial coordinate of the Boyer-Lindquist coordinate at infinity ends on spatial infinity rather than null infinity. On the other hand, it is known that the BMS symmetry is  the asymptotic symmetry of the asymptotically flat spacetimes at null infinity. Thus, performing holographic calculations in the Boyer- Lindquist coordinate requires extension of the BMS symmetry to spatial infinity.   It was argued in \cite{aspects},\cite{Compere:2014cna} that by defining a  modified bracket, the BMS  symmetry makes sense every where in the spacetime. Thus one can conclude that the BMS symmetry is also the asymptotic symmetry at spatial infinity of the asymptotically flat spacetimes. In fact the same result was established in \cite{Bagchi:2012cy} for spatial infinity by taking the flat limit from the asymptotic killing vectors of AdS spacetimes written in the global coordinate. In this paper we complete the calculation of \cite{Bagchi:2012cy} and propose appropriate boundary conditions at spatial infinity which result in generators of BMS symmetry.
 
 Using this symmetry, we can propose a holographic dual for the asymptotically flat spacetimes at spatial infinity. The correlators of the dual field theory are given by the contraction of a CFT. Thus we call it contracted CFT (CCFT)\footnote{In the original proposal of \cite{Bagchi:2010zz} this correspondence was named BMS/GCA. However, in four dimensions the symmetry algebra which is given by contraction of conformal algebra does not have Galilean subalgebra. Thus we will call the field theories which are dual of asymptotically flat spacetimes contracted conformal field theories.  }.    The correspondence between asymptotically flat spacetimes and CCFTs  can be used to propose a quasi local stress tensor for asymptotically flat spacetimes \cite{Fareghbal:2013ifa}. This proposal has also been checked for three dimensional Rindler space times \cite{Fareghbal:2014oba} , three dimensional hairy black holes \cite{Fareghbal:2014kfa} and the derivation of ultra-relativistic conformal anomaly in \cite{Fareghbal:2015bxd}.  The proposed stress tensor also generates the correct values for the conserved charges.  In this paper,  we also use the proposal of \cite{Fareghbal:2013ifa}  to find a stress tensor  for the Kerr black hole  in four dimensions. The starting point is the standard calculation of AdS/CFT for the quasi local  stress tensor of the Kerr-AdS written in the Boyer-Lindquist coordinate. We take the limit from this stress tensor and find a stress tensor that yields correct values of mass and angular momentum of the Kerr black hole .

  To find charges, we use the Brown and York's method \cite{Brown:1992br} which requires a timelike hypersurface. The charges are given by integrating over this timelike surface. For the Kerr-AdS this hypersurface is the boundary of the spacetime, but choosing this hypersurface  for the Kerr solution is challenging. We use a method that was previously tested in \cite{Fareghbal:2013ifa}-\cite{Fareghbal:2015bxd} and relate this hypersurface to the three dimensional spacetime in which the dual contracted conformal field theory lives on. It was proposed in \cite{Fareghbal:2013ifa} that the geometry of spacetimes, in which contracted CFT lives on, is the same as original CFT but with a contracted time coordinate. For the asymptotically flat spacetimes, this approach defines a conformal infinity that is given by anisotropic scaling of the metric components. This may have some roots in the connection between  the BMS symmetry as asymptotic symmetry of asymptotically flat spacetimes and the contraction of conformal symmetry. The analysis of  \cite{Horava:2009vy} might be helpful for further study of this problem.
  
  The fact that dual theory of four dimensional asymptotically flat gravity is a three dimensional theory with symmetries isomorphic to BMS$_4$ can be used for a direct calculation of energy momentum tensor components. In the gravity calculation of the stress tensor of the Kerr black hole we just take a limit from the Kerr-AdS stress tensor. However, in the field theory side we can use symmetries and directly calculate energy momentum tensor. The analogue calculation for the two dimensional CCFT or GCFT has been done in paper \cite{Bagchi:2015wna}. We perform a similar analysis and find some components of stress tensor. These components are completely consistent with the holographic result of the Kerr black hole.

Achieving flat-space holography, different proposals were introduced throughout the literature.
For instance, some of them suggest establishing a holographic dual at null infinity, the others propose a dual theory at  spatial infinity. If one insist on a holographic dual located at null infinity and defines the conformal boundary by using the standard method, the dual theory will be on a surface two dimensions less than gravitational theory. The main problem of this idea is the reconstruction of  the evolution of bulk fields  along two extra dimensions from the boundary field  theory data. This problem has been studied in  papers \cite{deBoer:2003vf}-\cite{Costa:2012fm} and some evidence has been provided which supports the conjecture that fields  on a specific class of asymptotically Ricci-flat spacetimes can be reconstructed out of conformal field theory data on a codimension two conformal manifold representing the boundary of a null surface in the bulk. However, the idea of defining conformal boundary by using anisotropic scaling can prevent  emergence of such problems. Moreover, the BMS symmetry in d+1 dimensions is isomorphic to the symmetry which is given by contraction of conformal symmetry in d dimensions. Thus the idea that dual theory of asymptotically flat spacetimes is on a spacetime with  just one dimension less, is more plausible.  Another way for  avoiding the technical problems of a dual description at null infinity is the proposal of holographic description  at spatial infinity \cite{Kraus:1999di}-\cite{Compere:2011db}. It is known that the standard holographic renormalization method for this case necessitates some non-local counter-terms for removing divergences. A holographic stress tensor for the Kerr black hole by using this direct method has been proposed in paper \cite{Astefanesei:2006zd}. However, an advantage of our method is that all of holographic data are constructed just by taking limit from the AdS/CFT calculations.

 In section two we review the results of \cite{Bagchi:2012cy} and introduce the generators of BMS$_4$ symmetry at the spatial infinity by proposing appropriate boundary conditions. In section three we   propose a stress tensor for the Kerr black hole by taking the limit of the Kerr-AdS case and observe that its conserved charges   have holographic description in terms of a field theory that is given by contraction of the CFT. In section four we performe a direct calculation of three dimensional CCFT energy momentum tensor and show the consistency between the  bulk and boundary calculations. The last section is devoted to the discussion and future directions.
 
\section{BMS$_4$ at Spatial infinity}
 The BMS group also acts at spatial infinity of the asymptotically flat four dimensional spacetimes. In order to find the generators of this symmetry we start with the isometries of $AdS_4$ in the  global coordinate 
\begin{equation}
 ds^2=-\left(1+{r^2\over\ell^2}\right)dt^2+{dr^2\over
  \left(1+{r^2\over\ell^2}\right)}+r^2\left(d\theta^2+\sin^2\theta d\phi^2\right),
 \end{equation} 
and take $\ell\to\infty$, which yields the Poincare symmetry. This limit has been taken in the appendix B of \cite{Bagchi:2012cy}. The final generators are 
\begin{eqnarray}\label{BMS4-gen1}
  \nonumber L_n&=& \frac12\left({xy-1\over xy+1}-n\right)x^n\left(t\p_t-r\p_r\)-x^{n+1}\p_x,\\
 \nonumber \bar L_n&=& \frac12\({xy-1\over xy+1}-n\)y^n\(t\p_t-r\p_r\)-y^{n+1}\p_y,\\
 M_{m,n}&=&{2\over 1+xy}x^m y^n \p_t,
  \end{eqnarray}  
 for $m,n=-1,0,1$ and 
 \begin{equation}\label{def of x y}
    x=e^{i\phi}\cot\frac\theta 2,\qquad y=e^{-i\phi}\cot\frac\theta 2 .
   \end{equation}     
 The generators   \eqref{BMS4-gen1} satisfy the following algebra,
 \begin{eqnarray}\label{BMS4-algebra}
 \nonumber [L_m,L_n]&=&(m-n)L_{m+n},\qquad [\bar L_m,\bar L_n]=(m-n)\bar L_{m+n},\qquad [L_m,\bar L_n]=0,\\
  {[}L_l,M_{m,n}{]}&=&\({l+1\over2}-m\)M_{m+l,n},\qquad   {[}\bar L_l,M_{m,n}{]}= \({l+1\over2}-n\)M_{m,n+l}. 
 \end{eqnarray}
This algebra can be given an infinite dimensional lift by defining $n$, $m$ and $l$ as any integer. We propose the generators \eqref{BMS4-gen1}  for any $n$ and $m$ as the generators of BMS$_4$ symmetry at    spatial infinity. We also introduce the appropriate boundary conditions which results in  \eqref{BMS4-gen1} as the asymptotic symmetry of four dimensional Minkowski spacetimes. In other words, we want to forget the flat-space limit and find proper boundary functions  which results in \eqref{BMS4-algebra}.To do so, we start with four dimensional Minkowski spacetime and write it in the following way:
\begin{equation}
 ds^2=g_{\mu\nu}^{ (0)}dx^\mu dx^\nu=-dt^2+dr^2+{4r^2\over (1+xy)^2}dx dy.
 \end{equation} 
We   define asymptotically flat spcatimes at spatial infinity by imposing boundary conditions 
\begin{align}\label{Boun.Cond.four.flat}
 \nonumber h_{tt}=\mathcal{O}(1),\qquad h_{tr}&=0,\qquad\qquad h_{tx}=\mathcal{O}(1), \qquad h_{ty}=\mathcal{O}(1),\\
\nonumber h_{rr}=\mathcal{O}(1),\qquad h_{rx}&=\mathcal{O}(r),\qquad h_{ry}=\mathcal{O}(r),\\
\nonumber h_{xx}=\mathcal{O}(1),\qquad  h_{yy}&=\mathcal{O}(1),\qquad h_{xy}=\mathcal{O}(1),\\
 h&=g^{\mu\nu(0)}h_{\mu\nu}=\mathcal{O}(1/r).
 \end{align}
 The asymptotic Killing vector which preserve \eqref{Boun.Cond.four.flat} is given by
 \begin{equation}\label{Four.Killing.vector}
 \xi=\left[T(x,y)+tf(x,y)\right]\p_t-rf(x,y)\p_r+X(x)\p_x+Y(y)\p_y,
 \end{equation}
 where $T(x,y)$, $X(x)$ and $Y(y)$ are arbitrary functions and
 \begin{equation}
 f(x,y)={\p_xX+\p_yY\over2}-{xY+yX\over 1+xy}.
 \end{equation}
 
Using \eqref{BMS4-gen1}, it is clear that 
\begin{align}
\nonumber L_n=\xi\left(T(x,y)=0, X(x)=-x^{n+1}, Y(y)=0 \), \\
\nonumber \bar L_n=\xi\left(T(x,y)=0, X(x)=0, Y(y)=-y^{n+1} \),\\
M_{m,n}=\xi\left(T(x,y)={2\over 1+xy}x^m y^n,X(x)=0,Y(y)=0\right).
 \end{align} 
 
 The algebra \eqref{BMS4-algebra} for $m,n,l={-1,0,1}$ can be generated by contraction of conformal algebra in three dimensions \cite{Alishahiha:2009np},\cite{Bagchi:2010zz},\cite{Bagchi:2012cy}. This is a good hint to find the corresponding operation of flat space limit in the boundary side. We propose it as a contraction of three dimensional  boundary CFT. To make this proposal more concrete we should contract n-point functions of operators in the boundary CFT. The simplest case is one point function of energy-momemntum tensor which corresponds to the quasi local stress tensor of the bulk geometry. We expect that the flat space limit of the stress tensor in the bulk side corresponds to the contraction of one point function of energy momentum tensor in the boundary theory. However, the flat space limit in the boundary side is not well-defined in all cases and we should use lessons of contraction of the boundary theory. In the next section we perform such a procedure for the simplest case which is Kerr black hole in the bulk side. We write the metric of the Kerr black hole in the Boyer-Lindquist coordinates which yields spatial infinity at $r\to\infty$. Since BMS$_4$ symmetry also acts at spatial infinity, the calculation of the next section make sense in the context of Flat/CCFT correspondence.

 \section{Holographic stress tensor and Flat/CCFT correspondence}
 \subsection{A brief review of three dimensional case}
 In order to find  the  stress tensor of the Kerr black hole, we need to use the method which was proposed in \cite{Fareghbal:2013ifa}. Hence   we briefly review  the results of \cite{Fareghbal:2013ifa} and then apply them for four dimensions. 
 We consider  three dimensional Einstein gravity without cosmological constant. A set of solutions of this theory can be written as 
 \begin{equation}\label{flat.BMS}
ds^{2} = M du^{2}-2dudr+2N dud\phi +r^{2}d\phi^{2},
\end{equation}
where
\bea\label{MN}
M = \theta(\phi),\qquad N = \chi(\phi)+\frac{u}{2}\theta^{\prime}(\phi),
\eea
$\theta$ and $\chi$ are arbitrary functions. The infinitesimal coordinate transformations which preserve \eqref{flat.BMS} are generated by the following vector fields:
\bea\label{xi.flat}
\xi^{r} = -r\partial_{\phi}Y +\partial_{\phi}^{2}F-\frac{1}{r}N\partial_{\phi}F,~~~~~~~\xi^{u} = F,~~~~~~~
\xi^{\phi} = Y -\frac{1}{r}\partial_{\phi}F,
\eea
where
\bea\label{BMS3}
Y = Y(\phi),~~~~~~~~~~~F = T(\phi)+u Y^{\prime}(\phi),
\eea
and $Y,T$ are arbitrary functions. At the leading order the generators
\bea\label{L,M}
L_{n} = \xi\big(Y = - e^{in\phi},T = 0\big),~~~~~~~~~M_{n} = \xi\big(Y = 0, T = -e^{in\phi}\big),
\eea
 form a representation of BMS$_3$ algebra on an asymptotically flat spacetimes
\bea\label{BMS.algebra}
[L_{m},L_{n}] = (m-n) L_{m+n},~~~~[L_{m},M_{n}] = (m-n) M_{m+n},~~~~
[M_{m},M_{n}] = 0.
\eea
It is not difficult to check that algebra \eqref{BMS.algebra} is given by Inonu-Wigner contraction of  two copies of the Virasoro algebra if we define
\bea\label{gravity.scaling}
L_{n} = \mathcal{L}_{n}-\bar{\mathcal{L}}_{-n},\qquad M_{n} = \epsilon(\mathcal{L}_{n}+\bar{\mathcal{L}}_{-n}),
\eea
and take $\epsilon\to 0$ limit. $\mathcal{L}_{n}$  and $\bar{\mathcal{L}}_{n}$ are generators of the Virasoro algebras and $\epsilon$ is a dimensionless parameter. The limit $\epsilon\to 0$ must be related to the flat-space limit. Thus we write it as $\epsilon={G\over \ell}$. Equation \eqref{gravity.scaling} is a hint which clarify how can we write the energy- momentum tensor of CCFT in terms of parent CFT energy-momentum tensor or the stress tensor of asymptotically flat spacetimes by taking limit from the asymptotically AdS counterpart in the gravity side.

The asymptotically flat metrics \eqref{flat.BMS} are given by taking flat-space limit from 
\bea\label{A.AdS.BMS}
ds^{2} = \left(-\frac{r^{2}}{l^{2}}+\mathcal{M}\right) du^{2} -2dudr+2\mathcal{N}dud\phi+r^{2}d\phi^{2},
\eea
where $\mathcal{M}$ and $\mathcal{N}$ are general functions of $u,\phi$ coordinates. Equations of motion resulted from Einstein  equations with negative cosmological constant are 
\bea
\partial_{u} \mathcal{M} =\frac{2}{l^{2}} \partial_{\phi}\mathcal{N},~~~~~~~2\partial_{u}\mathcal{N} =\partial_{\phi}\mathcal{M}.
\eea
In the coordinates $x^{+}, x^{-}$ which are defined as
\bea
x^{\pm} = \frac{u}{l} \pm \phi,
\eea
 $\mathcal{M},\mathcal{N}$  are given by
\bea\label{mMmN}
\mathcal{M} = 2\big(\Xi(x^{+})+\bar{\Xi}(x^{-})\big),~~~~~~~~\mathcal{N} = l\big(\Xi(x^{+})-\bar{\Xi}(x^{-})\big),
\eea
where $\Xi(x^{+})$ and $\bar{\Xi}(x^{-})$ are arbitrary functions of their argument. Moreover, 
\bea\
M =\lim_{G/\ell\to 0}\mathcal{M},\qquad N =\lim_{G/\ell\to 0}\mathcal{N}.
\eea
In order to find the stress tensor of the asymptotically flat spacetimes given by \eqref{flat.BMS}, we can calculate the stress tensor of the asymptotically AdS solutions \eqref{A.AdS.BMS} and take limit. According to \cite{Balasubramanian:1999re}-\cite{HH}, the quasi-local stress tensor is given by
  \begin{equation}\label{def.BY}
    T^{\mu\nu}={2\over\sqrt{-\gamma}}\dfrac{\delta S}{\delta \gamma_{\mu\nu}}, 
    \end{equation}  
 where $\gamma_{\mu\nu}=g_{\mu\nu}-n_\mu n_\nu$ is the boundary metric and $n_\mu$ is the outward pointing normal vector
to the boundary $\partial M$. $S$ is given by
\begin{equation}
S={1\over 16\pi G}\int\, d^3x \sqrt{-g}(R+{2\over\ell^2})-{1\over 8\pi G}\int_{\partial M} d^2x {\mathcal{K}}+S_{ct},
\end{equation}
 where  $\mathcal{K}$  is   extrinsic curvature of the boundary and $S_{ct}$ is added to remove divergent terms at the boundary. For this case we have
 \begin{equation}
 S_{ct}=-{1\over 8\pi G\ell}\int_{\partial M}  d^2x \sqrt{-\gamma}
 \end{equation}
The non-zero components of stress tensor at the boundary are
\begin{equation}\label{ST.AdS}
T_{uu}={\mathcal{M}\over 16\pi G\ell},\qquad T_{u\phi}={\mathcal{N}\over 8\pi G\ell},\qquad T_{\phi\phi}={\ell\mathcal{M}\over 16\pi G}.
\end{equation}
It is clear that taking the flat-space limit from \eqref{ST.AdS} is not well-defined. However, we make use of \eqref{gravity.scaling} and define the components of the stress tensor of three dimensional  asymptotically flat metrics,$\tilde T_{\mu\nu}$ by 
\begin{eqnarray}\label{def of energy-momentum }
 \nonumber \tilde T_{++}+\tilde T_{--}&=&\lim_{\epsilon\to 0}\epsilon\left(T_{++}+T_{--}\right)\\
  \nonumber\tilde T_{++}-\tilde T_{--}&=&\lim_{\epsilon\to 0}\left(T_{++}-T_{--}\right)\\
  \tilde T_{+-}&=&\lim_{\epsilon\to 0} \epsilon\, T_{+-}.
  \end{eqnarray}  
where light-cone coordinates for the asymptotically flat case are given by ${u\over G}\pm \phi $. Using \eqref{def of energy-momentum } we find 
\bea\label{flat.stress.tensor}
\tilde{T}_{uu} =\frac{M}{16\pi G^{2}},\qquad\tilde{T}_{u\phi} = \frac{N}{8\pi G^{2}},\qquad\tilde{T}_{\phi\phi} = \frac{M}{16\pi}.
\eea
Comparison of \eqref{ST.AdS} and \eqref{flat.stress.tensor} is instructive. No doubt that \eqref{def of energy-momentum } has roots in the dual filed theory (see for example \cite{Bagchi:2010vw}) but \eqref{ST.AdS} and \eqref{flat.stress.tensor} show that in the gravity side one can scale the components of stress tensor by some appropriate powers of $\ell$ and then take the flat-space limit. We will use this idea in the next subsection and propose a stress tensor for the Kerr black hole. We should note  that  a deep relation between the scaling of components in the gravity side and a definition similar to \eqref{def of energy-momentum } which is originated from the dual field theory is also expectable for the Kerr black hole. However, since a three dimensional field theory with $BMS_4$ symmetry has not been studied very carefully, a relation similar to \eqref{def of energy-momentum } is not clear for us in this stage.

\subsection{A quasi local stress tensor for the Kerr black hole } 
 In this section we want to propose a quasi-local stress tensor for the Kerr black hole. We start with the Kerr-AdS black hole which is given by
 \begin{equation}\label{kerr-AdS}
 ds^2=-{\Delta_r\over\rho^2}\left(dt-{a \sin^2\theta\over \Xi}d\phi\right)^2+{\rho^2\over \Delta_r}dr^2+{\rho^2\over \Delta_\theta}d\theta^2+{\Delta_\theta \sin^2\theta\over\rho^2}\left(a dt-{r^2+a^2\over\Xi}d\phi\right)^2,
 \end{equation}
 where
 \begin{align}
 &\Delta_r=(r^2+a^2)\left(1+{r^2\over\ell^2}\right)-2M G r,\qquad \Delta_\theta=1-{a^2\over\ell^2}\cos^2\theta,\\
 &\rho^2=r^2+a^2\cos^2\theta,\qquad \Xi=1-\dfrac{a^2}{\ell^2}.
 \end{align}
 $M$ is the mass of the black hole and $a=J/M$ where $J$ is the angular momentum of the black hole. This black hole is asymptotically AdS  with radius $\ell$. The line element \eqref{kerr-AdS} is the solution of the following theory:
 \begin{equation}\label{action-0}
  S_0=\dfrac{1}{16\pi G}\int d^4x \sqrt{-g}\left(R+{6\over\ell^2}\right).
  \end{equation}
  We want to calculate the quasi-local stress tensor of the Kerr-AdS \eqref{kerr-AdS} by using Brown and York's method \cite{Brown:1992br}. According to the dictionary of the AdS/CFT correspondence, the components of this tensor correspond to the expectation values of the energy-momentum tensor of the boundary CFT.  The quasi-local stress tensor is given again  by \eqref {def.BY}.   $S$ is given by
\begin{equation}
S=S_0-{1\over 8\pi G}\int_{\partial M} d^3x {\mathcal{K}}+S_{ct},
\end{equation}
 where $S_{ct}$ is 
 \begin{equation}
 S_{ct}=-{1\over 4 \pi G \ell}\int_{\partial M} d^3 x \sqrt{-\gamma}\left(1-{\ell^2\over 4}R_{(3)}\right)
 \end{equation}
 where $R_{(3)}$ is the Ricci scalar of $\gamma_{\mu\nu}$. Calculation of the quasi-local stress tensor for the Kerr-AdS, $T_{ij}$, has been done in papers \cite{Caldarelli:1999xj}-\cite{Papadimitriou:2005ii} and the non-zero components (up to $1/r$ order) are given by
 \begin{align}\label{Kerr-AdS-first}
\nonumber 8\pi T_{tt}&={2M\over r\ell},\\
\nonumber 8\pi T_{t\phi}&=-{2aM\over r\ell\Xi}\sin^2\theta,\\
\nonumber 8\pi T_{\theta\theta}&={M\ell\over r\Delta_\theta},\\
 8\pi T_{\phi\phi}&={M\ell\over r\Xi^2}\sin^2\theta\left(\Xi+{3a^2\sin^2\theta\over\ell^2}\right).
 \end{align}
One can use \eqref{Kerr-AdS-first} and compute the conserved charges of Kerr-AdS black hole. This method was introduced by Brown and York \cite{Brown:1992br} which defines charge $Q_\xi$ associated to symmetry generator $\xi$ as
\begin{equation}\label{def-of-charge}
Q_\xi=\int_\Sigma d\sigma \sqrt{\sigma}v^\mu\xi^\nu T_{\mu\nu},
\end{equation}
 where $\Sigma$ is a $t$-constant surface in $\partial M$,$\sigma_{ab}$ is metric of $\Sigma$ and $v^\mu$ is the unit timelike vector normal to $\Sigma$. For the Kerr-AdS black hole given by \eqref{kerr-AdS}, the metric of the boundary $\partial M$ is given by 
 \begin{equation}\label{boundary-first}
 ds^2={r^2\over\ell^2}\left(-dt^2+{2a\sin^2\theta\over\Xi}dt d\phi+{\ell^2\over\Delta_\theta}d\theta^2+{\ell^2\sin^2\theta\over\Xi}d\phi^2\right).
 \end{equation}
 It is clear that \eqref{boundary-first} does not describe a $R\times S^2$ boundary. In order to change the metric of $\partial M$ to the standard form of the boundary of the global AdS we use of a coordinate transformation
 \begin{equation}\label{coord-change}
 \phi=\varphi-{a\over\ell^2}t.
 \end{equation}
This transformation yields,
 \begin{align}\label{EMT after coord-change}
\nonumber 8\pi T_{tt}&={2M\over r\ell}+\mathcal{O}({1\over\ell^3}),\\
8\pi T_{t\varphi}&=-{3aM\over r\ell\Xi}\sin^2\theta+\mathcal{O}({1\over\ell^3}).
 \end{align}
    $T_{\varphi\varphi}$ is the same as $T_{\phi\phi}$ and the other components remain unchanged.  
 
 Now we want to use \eqref{Kerr-AdS-first} and \eqref{EMT after coord-change} in order to take the flat-space limit to introduce a quasi local stress tensor for the Kerr black hole. It is clear from \eqref{Kerr-AdS-first} and \eqref{EMT after coord-change} that for $\ell\to\infty$ some of the components of the stress tensor are zero and some of them diverge. Therefore at first look, using this approach, while defining flat spacetime stress tensor by taking flat space limit, is not well-defined. However, there is a proposal in \cite{Fareghbal:2013ifa} that tackles this problem and results in well-defined components for  the stress tensor of the asymptotically flat spacetimes. According to \cite{Fareghbal:2013ifa} if one accepts the correspondence of the flat space limit in the bulk as the contraction of the CFT at the boundary, it is possible to use one point functions of the CCFT energy momentum tensor.There also exists a method to find a correct definition of flat space stress tensor. This method  has been established in \cite{Fareghbal:2013ifa} for three dimensions. The final result was interesting and its extension to higher dimensions is plausible. The lessons of \cite{Fareghbal:2013ifa} show that in order to find the flat space stress tensor from the AdS counterparts, one must apply proper powers of $\ell$ to the components of the stress tensor, so that the flat limit is well defined. This factors must be dimensionless, so we construct them in four dimensions by using $\ell/\sqrt{G}$. Keeping this proposal in mind and using \eqref{Kerr-AdS-first} and \eqref{EMT after coord-change} we introduce the Kerr black hole stress tensor, $\tau_{ij}$  as 
 \begin{align}\label{Kerr stress tensor}
 \nonumber 8\pi \tau_{tt}&={2M\over r\sqrt{G}},\\
\nonumber 8\pi \tau_{t\varphi}&=-{3aM\over r\sqrt{G}}\sin^2\theta,\\
\nonumber 8\pi \tau_{\theta\theta}&={M\sqrt{G}\over r},\\
 8\pi \tau_{\varphi\varphi}&={M\sqrt{G}\over r}\sin^2\theta.
 \end{align}
 
 We can use \eqref{Kerr stress tensor} and calculate conserved charges of Kerr black hole by using \eqref{def-of-charge}. However, in this case the hypersurface $\partial M$ is not the boundary and it is necessary that we determine this hypersurface and its t-constant surface $\Sigma$ carefully. We found \eqref{Kerr stress tensor} by taking limit and we also expect to find $\Sigma$  by taking limit from the Kerr-AdS counterpart. The proposal has also been established in \cite{Fareghbal:2013ifa} for the three dimensioanl asymptotically flat solutions. We plan to use this proposal for the current problem. The key point is if we assume the flat space limit of the bulk theory as a contraction of the boundary theory, the metric of spacetimes, which the boundary  theories lives on , are related  before and after contraction.  They are the same with a difference that one of the coordinates of the spacetime, which contracted theory lives on, is  contracted coordinate of the original theory before contraction. To be precise, let us use \eqref{boundary-first} and try to understand the role of $\ell$  in the definition of the conformal boundary. According to the AdS/CFT correspondence, the conformal boundary is used for finding the spacetime  in which the dual CFT lives on . After applying coordinate change \eqref{coord-change}, the boundary metric \eqref{boundary-first} is written as
\begin{equation}\label{boundary2}
 ds^2={r^2\over G}\left[-\left(1+{a^2\sin^2\theta\over \Xi\ell^2}\right){G\over\ell^2}dt^2+{G\over\Delta_\theta}d\theta^2+{G\sin^2\theta\over\Xi}d\varphi^2\right],
 \end{equation}   
 where we have used Newton constant $G$ to make the conformal factor dimensionless. If we  define a new time $\tilde t$ as
 \begin{equation}\label{cont time}
 \tilde t={\sqrt{G}\over \ell}t,
 \end{equation}
 the flat limit, $\ell\to\infty$ is well-defined and yields
 \begin{equation}\label{boundaryflat}
 d\tilde s^2={r^2\over G}\left[-d\tilde t^2+G d\theta^2+{G\sin^2\theta}d\varphi^2\right].
 \end{equation}
It is clear from  \eqref{cont time} that $\tilde t$ only make sense for finite $\ell$, but as $\ell\to\infty$ it is nothing other than the contraction of time for the boundary theory. We assume that the dual theory of asymptotically flat spacetimes lives on a spacetime where its time is given by contraction of the time coordinate of original CFT. In other words the dual of Kerr black hole must live on a spacetime with a metric given by \eqref{boundaryflat} and $\tilde t$ is the time coordinate of the Kerr. Returning $\tilde t$ to $t$ and defining $\partial M$ in \eqref{def-of-charge} as \eqref{boundaryflat}, we can use \eqref{Kerr stress tensor} and calculate the conserved charges of Kerr black hole. It is not difficult to check that 
\begin{equation}
Q_{\partial t}=M,\qquad Q_{\partial \varphi}=-aM, 
\end{equation}
are the mass and angular momentum of the Kerr black hole. 
 
 \section{A field theoretic approach}
 
 Our calculation in the previous section was on the gravity side and we just took limit from the stress tensor of the Kerr-AdS black hole. The stress tensor of the bulk theory corresponds to the energy-momentum tensor of the boundary theory. Thus we have a chance for checking the correctness of  the calculation of the previous section by directly calculating the energy momentum tensor of the three dimensional CCFT.
 
 Making the discussion relevant with the former section, let us consider a three dimensional CCFT on a flat manifold with metric
  \begin{equation}\label{3d flat manifold}
 d\tilde s^2={r^2\over G}\left[-d t^2+G d\theta^2+{G\sin^2\theta}d\varphi^2\right]={r^2\over G}\left[-d t^2+{4G\,dx\,dy\over (1+xy)^2} \right].
 \end{equation} 
 where $x$ and $y$ are defined by \eqref{def of x y} and $r$ is just a conformal factor. The symmetry algebra of CCFT is isomorphic to the  BMS$_4$ algebra  and is given by \eqref{BMS4-algebra}. The symmetry generators are 
\begin{eqnarray}\label{3d ccft symmetry}
  \nonumber L_n&=& \frac12\left({xy-1\over xy+1}-n\right)x^n t\p_t-x^{n+1}\p_x,\\
 \nonumber \bar L_n&=& \frac12\({xy-1\over xy+1}-n\)y^n t\p_t-y^{n+1}\p_y,\\
 M_{m,n}&=&{2\over 1+xy}x^m y^n \p_t.
  \end{eqnarray} 
 
 The starting point is  definition of the conserved charges in the field theory,
 \begin{equation}\label{def of charge CCFT}
 Q={r^3\over\sqrt{G}}\int d\theta\, d\phi\, \sin \theta\, J^t={r^3\over\sqrt{G}}\int d\theta\, d\phi\, \sin \theta\, T^{t\mu}\xi_\nu
 \end{equation}
 where $J^\mu$ is the corresponding current of symmetries, $T^{\mu\nu}$ is the energy-momentum tensor and $\xi^\mu$ is the generator of  symmetry. Using \eqref{3d flat manifold} and \eqref{3d ccft symmetry} we have
 \begin{equation}
 Q_{M_{m,n}}={2r^5\over G\sqrt G}\int d\theta\,d\phi\, \sin\theta {x^m\,y^n\over 1+xy}T^{tt}. 
 \end{equation}

 Using
 \begin{eqnarray}\label{delta rel.}
 \nonumber \sum_{m} x^{m-\frac{1}{2}}x'^{-m-\frac{1}{2}}=2\pi i \delta(x-x'),\\
  \sum_{m} y^{m-\frac{1}{2}}y'^{-m-\frac{1}{2}}=2\pi i \delta(y-y'),
 \end{eqnarray}
 we can simplify the above equation and write 
 \begin{equation}\label{Ttt CCFT}
 T^{tt}={G\sqrt G(1+xy)^3\over 16\pi^3i r^5 }\sum_m\sum_n Q_{M_{m,n}}x^{-m-1} y^{-n-1}.
 \end{equation}
 This result determines $T^{tt}$ in CCFT and the corresponding calculation  is a pure field theoretic one. Now we want to relate this component of the CCFT energy momentum tensor to the stress tensor of the bulk theory.  Thus we should calculate $Q_{M_{m,n}}$ in the bulk side. In particular, we are interested in the case that the bulk solution is the Kerr black hole. In the gravity side $M_{m,n}$  are the generators of asymptotic symmetry, it is possible to find the corresponding charges of these symmetries by various methods. The most applicable method is  covariant  formalism developed in \cite{Barnich:2001jy}. The charges associated to $M_{m,n}$  for the Kerr black hole were calculated using this method in paper \cite{Barnich:2011mi}\footnote{In our convention  $M$ is the mass of the black hole and it is different from $M$ in paper \cite{Barnich:2011mi} by a factor of $G$.},
 \begin{equation}\label{Mmn charge Kerr}
 Q_{M_{m,n}}={M\over 2\pi }\int d^2\Omega {x^my^n\over(1+xy)}
 \end{equation}
If we substitute \eqref{Mmn charge Kerr} in \eqref{Ttt CCFT} and use \eqref{delta rel.} we will find a result which is in agreement with \eqref{Kerr stress tensor}. Moreover, using \eqref{3d flat manifold}, \eqref{3d ccft symmetry} and  \eqref{def of charge CCFT} we can write
\begin{equation}\label{def of Ln charge}
 Q_{L_n}={r^5\over\sqrt G}\int d\theta d\phi \sin\theta\left[\frac 12 t x^n {T^{tt}\over G}\left({xy-1\over xy+1}-n\right)+{2x^{n+1} T^{ty}\over (1+xy)^2} \right],
\end{equation}
\begin{equation}\label{def of bLn charge}
Q_{\bar L_n}={r^5\over\sqrt G}\int d\theta d\phi \sin\theta\left[\frac 12 t y^n {T^{tt}\over G}\left({xy-1\over xy+1}-n\right)+{2y^{n+1} T^{tx}\over (1+xy)^2} \right]. 
\end{equation}
 The above equations  are  again written in the field theory side. We can find the corresponding elements of energy momentum tensor as well as charges in the gravity side. The calculation of charges for the Kerr black hole results in\cite{Barnich:2011mi}
\begin{equation}\label{qln qbln}
Q_{L_n}=-Q_{\bar L_n}=-i{aM\over 2}\delta_n^0.
\end{equation}
 If we substitute $T^{tt}$, $T^{tx}$ and $T^{ty}$ from \eqref{Kerr stress tensor}, the right hand side of \eqref{def of Ln charge} and
 \eqref{def of bLn charge} yields exactly \eqref{qln qbln}. This is a non-trivial check that our stress tensor for the Kerr black hole \eqref{Kerr stress tensor} which is given just by taking flat space limit, is meaningful if one uses the Flat/CCFT correspondence.

\section{Discussion }
In this paper, the proposal of \cite{Fareghbal:2013ifa} has been investigated in four dimensions. The main results of \cite{Fareghbal:2013ifa} are based on the correspondence between asymptotically flat space times and contracted CFTs. Thus, in this view we started establishing a holographic picture for  four dimensional asymptotically flat spacetimes. The main cavity for the generalizing method of \cite{Fareghbal:2013ifa} was that defining four dimensional  asymptotically flat spacetimes in the BMS gauge is not appropriate when one wants to use the flat-space limit technique. However, the work implemented in this paper argues that using another coordinates system can solve this problem. Demonstrated in the first example, we used the Kerr-AdS black hole written in the Boyer-Lindquist coordinate. Using this coordinate and taking the flat space limit requires the dual  boundary theory of  Kerr to live on spatial infinity rather than null infinity. This requirement necessitates an asymptotic symmetry group at spatial infinity. We proposed that this symmetry is still BMS symmetry and introduced the boundary condition and the symmetry generators.

The calculations done in this paper represent the first step to accomplish the aforementioned. The next step will be generalizing this method for arbitrary asymptotically flat spacetimes in four dimensions. The boundary conditions, which these spacetimes must satisfy at spatial infinity, were introduced in this paper. Nevertheless the generic solution to the field equations, which obeys  these boundary conditions, must be explored. Using the generic solution one can calculate the charges of the asymptotic killing vectors and see whether they are finite, conserved and integrable. 

\subsubsection*{Acknowledgements}
The authors would like to  thank Shannon Ray for his comments on the manuscript. We are also grateful to Ali Naseh and Seyed Morteza Hosseini for useful comments and discussions.

\appendix


\end{document}